\documentclass[12pt]{iopart}

\usepackage{graphicx}
\begin{document}

\title{Orbital-selective Mott phase of Cu-substituted iron-based superconductors}

\author{Yang Liu, Yang-Yang Zhao, and Yun Song}
\address{Department of Physics, Beijing Normal University, Beijing 100875, China}
\ead{yunsong@bnu.edu.cn}

\date{\today}

\begin{abstract}
We study the phase transition in Cu-substituted iron-based
superconductors with a new developed real-space Green's function
method. We find that Cu substitution has strong effect on the
orbital-selective Mott transition introduced by the Hund's rule
coupling. The redistribution of the orbital occupancy which is
caused by the increase of the Hund's rule coupling, gives rise
to the Mott-Hubbard metal-insulator transition in the half-filled
$d_{xy}$ orbital. We also find that more and more electronic states
appear inside that Mott gap of the $d_{xy}$ orbital with the increase
of Cu substitution, and the in-gap states around the Fermi level
are strongly localized at some specific lattice sites. Further,
a distinctive phase diagram, obtained for the Cu-substituted
Fe-based superconductors, displays an orbital-selective
insulating phase, as a result of the cooperative effect of the
Hund's rule coupling and the impurity-induced disorder.
\end{abstract}


\maketitle


\section{INTRODUCTION}
\label{sec:INTR}

The multiorbital nature of iron pnictides and chalcogenides
is essential for understanding the mechanism of the
high-temperature superconductivity of iron-based superconductors
\cite{Lebegue2007,Singh2008}. The multiple Fe 3$d$ orbitals,
which couple strongly with each other by the Hund's rule exchange
interaction, contribute most to the low-energy excitations
\cite{Hosono2006,Hirschfeld2011,ChenX2014}.
Some theoretical studies showed that the Hund's rule coupling
plays a key role in tuning the degree of electronic correlation
in multiorbital systems
\cite{Haule2009,Mecici2011,Medici2014,Fanfarillo2015}. Using a slave-spin method, it was
predicted that an orbital-selective Mott phase (OSMP) will emerge
in K$_{1-x}$Fe$_{2-y}$Se$_2$ when the Hund's rule coupling exceeds
a certain threshold value \cite{YuR2013}. Experimentally, strong
orbital-dependence of the correlation was found in iron
chalcogenides \cite{YiM2015}, and the signs of the appearance of OSMP
were also observed in A$_x$Fe$_{2-y}$Se$_2$ (A=K, Rb) at high temperature
\cite{YiM2013,WangZ2014,DingXX2014}.

It has been demonstrated by the dynamical mean-field theory (DMFT)
\cite{Rev-DMFT} that, in a multiorbital system, the OSMP is
usually derived from
three possible reasons: (1) the bandwidths of the two orbitals are
significantly different \cite{Anisimov2002}; (2) the influence of
the crystal-field splitting is considerable \cite{Werner2007}; and
(3) the orbital degeneracy is reduced owing to the distinct
features of the orbitals or by some other reasons
\cite{Medici2009,SongZ2014}. However, it is still unclear whether
the OSMP is sensitive to the effect of disorder introduced by
impurities. In iron-based superconductors, many experiments were
implemented by substituting Fe with other transition metals
\cite{McLeod,YanYJ,ChengP,Ideta,Merz,WangAF,CuiST,HuangTW,NiN,LiJ,Williams},
which not only affects the carrier density of the system, but also
introduces the effect of disorder. Among various transition-metal
substitutions, Cu doping is highly disruptive to the electron
structures of the FeAs sheets. In this paper, we will focus on the
effect of Cu substitution in Fe-based superconductors, especially
the cooperative effect of the Hund's rule coupling and the
impurity-induced disorder on the OSMP.

It is still a theoretical challenge to study the multiorbital
correlation and the effect of disorder on equal footing
\cite{AL50years}. To precisely study the real-space
fluctuations introduced by Cu substitution, we develop a
real-space Green's function method by extending the Hubbard-I
approximation \cite{Hubbard-I} to the multiorbital Hubbard model.
Our decoupling approximation has been proved to be adequate for the
study of Cu-substituted iron-based superconductors \cite{Yangliu}.
In according with the prediction of other theoretical studies
\cite{Mecici2011,Medici2014,YuR2013}, we find an OSMP tuned by the Hund's rule
coupling in the undoped cases, where the $d_{xy}$ orbital is a Mott
insulator while the $d_{xz}/d_{yz}$ orbitals remain itinerant. In
the presence of the Cu substitution, the competition between the
Mott-Hubbard and Anderson metal-insulator transitions (MIT) is
reveled by the study of the interplay between the Hund's rule
coupling and the impurity-induced disorder in Cu-substituted
Fe-based superconductors. We find that the Cu substitutions can
introduce some states inside the Mott gap of the $d_{xy}$ orbital
around the Fermi level, which can be suppressed by the increasing
Hund's rule coupling. As a result, the increase of the critical value
for the Mott-Hubbard MIT transition is found in the $d_{xy}$ orbital.
Meanwhile, the degenerate $d_{xz}/d_{yz}$ orbitals are Anderson
localized, indicating the existence of an orbital-selective insulating
phase in Fe-based superconductors with the presence of impurities. We
also construct a complete phase diagram for Cu-substituted iron-based
superconductors, depended on both the Hund's rule coupling and the
concentration of Cu impurities.

The paper is organized as follows. In section~\ref{Sec:Model}, we
introduce an inhomogeneous three-orbital Hubbard model, and develop
a real-space Green's function approach to study the cooperative
effect of disorder and multiorbital correlation in Cu-substituted Fe-based
superconductors. In section~\ref{Sec:U+J}, we first show the
orbital-selective Mott transition (OSMT) introduced by the Hund's
rule coupling in undoped compound. Then we study the cooperative
effect of the Hund's rule coupling and impurity-induced disorder
on the phase transitions in Cu-substituted iron-based superconductors.
Further, a phase diagram is obtained by the study of the competition
between the Mott-Hubbard and Anderson MITs. The principal findings of
this paper are summarized in section~\ref{Sec:Con}.


\section{Model Hamiltonian and Methodology}
\label{Sec:Model}


An inhomogeneous three-orbital Hubbard model is introduced for the
Cu-substituted Fe-based compounds based on the local density
functional calculations of NaFe$_{1-x}$Cu$_x$As \cite{Yangliu},
where a certain percentage of Fe sites is replaced by Cu ions
randomly. The Hamiltonian is expressed as
\begin{eqnarray}
\textrm{H}&=&-\sum_{i\neq j (i,j\in Fe)}\sum_{\alpha\beta\sigma}
                  T_{i\alpha,j\beta}c_{i\alpha\sigma}^\dag c_{j\beta\sigma}
              -\sum_{i\neq j (i,j\in Cu)}\sum_{\alpha\sigma}
                  t_{ij}c_{i\alpha\sigma}^\dag c_{j\alpha\sigma}\nonumber\\
          &&-\sum_{i\in Fe,j\in Cu}\sum_{\alpha\beta\sigma}
                  t_{i j}^{\prime} c_{i\alpha\sigma}^\dag c_{j\beta\sigma}
             -\mu\sum_{i\alpha\sigma}n_{i\alpha\sigma}+\Delta_{xy}\sum_{i\in Fe, \sigma}
              n_{i,xy, \sigma}\nonumber\\
          &&+U\sum_{i\in Fe,\alpha}n_{i\alpha\uparrow}n_{i\alpha\downarrow}
              +\sum_{i\in Fe}\sum_{\alpha<\beta,\sigma\sigma'}(U'-J \delta_{\sigma\sigma'})
              n_{i\alpha\sigma}n_{i\beta\sigma'}\nonumber\\
          &&+u\sum_{i\in Cu,\alpha} n_{i\alpha\uparrow}n_{i\alpha\downarrow},
 \label{Eq_Ham}
\end{eqnarray}
where $c_{i\alpha\sigma}^\dag$ ($c_{i\alpha\sigma}$) creates
(annihilates) an electron with spin projection $\sigma$ for orbital
$\alpha$ ($d_{xz}$, $d_{yz}$, or $d_{xy}$) of an iron or copper site
$i$. $T_{i\alpha,j\beta}$ is the hopping term within orbitals
$\alpha$ and $\beta$ between nearest-neighbor (NN) or
next-nearest-neighbor (NNN) iron sites $i$ and $j$. $t_{ij}$
represents the intraorbital hopping integral between NN copper
atoms, and $t_{ij}^{\prime}$ denotes the hopping between NN Cu and
Fe ions. $U$ and $U^{\prime}$ are the onsite intraorbital and
interorbital Coulomb interactions on iron sites respectively, and
$\Delta_{xy}=\mu_{xy}-\mu_{xz}$ is the energy splitting between the
$d_{xy}$ and the degenerate $d_{xz}$/$d_{yz}$ orbitals. Further,
$J$ represents the Ising-type Hund's rule couplings for Fe atoms,
and $u$ is the intraorbital interaction of Cu ions.

\begin{table*}[htb]
\begin{center}
\caption{Every single term of the nearest neighbor
and next nearest neighbor hoppings ($T_{NN}$ and
$T_{NNN}$) for the Fe ions of the tight-binding
Hamiltonian in Eq.~(\ref{Eq_Ham}). $(-1)^{|i|}$
indicates that the hopping parameters change sign
along the site locations, arising from the two-iron
unit cell of the original FeAs planes \cite{Daghofer2010}.} \

\begin{tabular}{c|rrrrr}
\hline \hline \\ [-4pt]
  $T_{NN}$
 & \big[$1,0$\big]
 & \big[$0,1$\big]
 & \big[$-1,0$\big]
 & \big[$0,-1$\big]
\\ [+4pt]
 \hline \\ [-5pt]             
 $(d_{xy},d_{xy})$                    &              $T_1$ &              $T_1$ &                $T_1$ &                $T_1$ &                 \\ [+4pt]
 $(d_{xy},d_{xz})$                    &  ~~$(-1)^{|i|}T_3$ &                  0 &  ~~$(-1)^{|i+1|}T_3$ &                    0 &                 \\ [+4pt]
 $(d_{xy},d_{yz})$                    &                  0 &  ~~$(-1)^{|i|}T_3$ &                    0 &  ~~$(-1)^{|i+1|}T_3$ &                 \\ [+4pt]
 $(d_{xz},d_{xz})$                    &              $T_6$ &              $T_5$ &                $T_6$ &                $T_5$ &                 \\ 
 $(d_{yz},d_{yz})$                    &              $T_5$ &              $T_6$ &                $T_5$ &                $T_6$ &                 \\ [+4pt]
 $(d_{yz},d_{xz})$                    &                  0 &                  0 &                    0 &                    0 &                 \\ [+4pt]
\hline \\ [-4pt]
  $T_{NNN}$
 & \big[$1,1$\big]
 & \big[$-1,1$\big]
 & \big[$-1,-1$\big]
 & \big[$1,-1$\big]
\\ [+4pt]
 \hline \\ [-5pt]             
 $(d_{xy},d_{xy})$                    &              $T_2$ &                $T_2$ &              $T_2$ &                $T_2$ &  \\ [+4pt]
 $(d_{xy},d_{xz})$                    &    ~~$(-1)^{|i+1|}T_4$ &    ~~$(-1)^{|i|}T_4$ &  ~~$(-1)^{|i|}T_4$ &  ~~$(-1)^{|i+1|}T_4$ &  \\ [+4pt]
 $(d_{xy},d_{yz})$                    &                    ~~$(-1)^{|i+1|}T_4$ &  ~~$(-1)^{|i+1|}T_4$ &  ~~$(-1)^{|i|}T_4$ &    ~~$(-1)^{|i|}T_4$ &  \\ [+4pt]
 $(d_{xz},d_{xz})$                    &                              $T_7$ &                $T_7$ &              $T_7$ &                $T_7$ &  \\ 
 $(d_{yz},d_{yz})$                    &                            $T_7$ &                $T_7$ &              $T_7$ &                $T_7$ &  \\ [+4pt]
 $(d_{yz},d_{xz})$                    &                                 $-T_8$ &                $T_8$ &             $-T_8$ &                $T_8$ &  \\ [+4pt]
\hline \hline
\end{tabular}
\end{center}
\label{Tb:Hop}
\end{table*}

The values of the tight-binding (TB) model parameters were determined by the first
principle calculations of the undoped Fe-based compounds
\cite{Daghofer2010,Daghofer2012}.
The local density function calculation predicted that the Fermi
surfaces of NaFe$_{1-x}$Cu$_x$As are mainly composed by three Fe
$3d$ orbitals, i.e. $d_{xy}$, $d_{xz}$ and $d_{yz}$ orbitals,
whereas Cu $3d$ orbitals distribute from -4 eV to -2 eV, which is far below
the Fermi level \cite{Yangliu}. Therefore, the Fermi surfaces should
be very sensitive to the hopping terms within the three Fe $3d$
orbitals. In the undoped case, our TB model for iron
sites is exactly the same as the three-orbital TB Hamiltonian
introduced by Daghofer $et$ $al.$ for the pnictides followed the
Slater-Koster procedure \cite{Daghofer2010,Daghofer2012}.
In Table~1, we display the details about the definitions of the
intraorbital and interorbital hoping integrals $T_{i\alpha,j\beta}$
for $d_{xy}$, $d_{xz}$ and $d_{yz}$ orbitals of Fe ions. Further
details about the model parameters can also be found in
Ref.~\cite{Yangliu}.

Although the correlations in various iron-based compounds are
different, it is generally believed that the correlations should
be smaller than the bandwidths. The parameters of the intraorbital
and interorbital interactions of the three-orbital Hubbard model
(Eq.~(\ref{Eq_Ham})) have been obtained
by the best fitting of the Fermi surfaces with the experimental
result of LiFeAs \cite{YiM2012}, that is $U$=1.5 eV and $U'$=1.125
eV \cite{Yangliu}. However, some study also suggested that the
Hund's rule coupling in iron-based superconductors should be within
$U/10$ and $U/5$ \cite{Haule2009}. Therefore, it is important to
find out what is the influence of different values of $J$ on the
properties of Fe-based superconductors. Because the main purpose of
this paper is to explore the cooperative effects of the Hund's rule
coupling and the doping-induced disorder, we fix the value of $U$ as
1.5 eV in the following calculations. The values of the TB model
parameters and the multiorbital correlations of the three-orbital
Hubbard model are shown in Table~\ref{Tb:pmts}.

\begin{table}[htbp]
\begin{center}
\caption{Values of the model parameters of the inhomogenous
three-orbital Hubbard model. The energy unit is electron volts. }\

\begin{tabular}{p{1.2cm}p{1.2cm}p{1.2cm}p{1.2cm}p{1.5cm}p{1.5cm}p{1.5cm}p{1.2cm}}
\hline \hline  \\ [-3pt]
 $  T_1$  & $T_2$ & $T_3$          &  $T_4$  & $T_5$        & $T_6$       & $T_7$     & $T_8$ \\
 \hline \\ [-3pt]
 0.15     & 0.15  & $-$0.12        &  $0.06$ & $-$0.08      & 0.1825      & 0.08375   & $-$0.03          \\
\hline\hline
$  t$     & $t'$  & $\Delta_{xy}$  &  $U$    & $U'$         & $J$         & $u$       &    \\
 \hline \\ [-3pt]
 0.01     & 0.01  &  0.75          &  1.5    & 0.9$\sim$1.5 & 0$\sim$0.3  & 3 or 6    &          \\
\hline\hline
\end{tabular}
\end{center}
\label{Tb:pmts}
\end{table}

We study the inhomogeneous three-orbital Hubbard model by a newly
developed real-space Green's function method \cite{Yangliu}. In this
approach, we generalize hundreds of disorder configurations for each
Cu concentration, and the final physical properties, such as the
density of states (DOS) and conductivity, are obtained by the
arithmetical average of the results for different samples. For a
certain disorder sample of a square lattice with $N=L^2$ sites, we
randomly choose $N_{imp}$ ($N_{imp}=N\times x$) sites to put the Cu
ions  when the concentration of Cu substitution is $x$. To exactly
study the effect of disorder, we construct a complete real-space
Green's function $\textbf{G}$, which can be expressed as a $3N\times
3N$ matrix
with elements defined as
\begin{eqnarray}
G_{ij\sigma}^{\alpha\beta}&=&\langle\langle c_{i\alpha\sigma} \mid
c^{\dag}_{j\beta\sigma}\rangle\rangle.
\end{eqnarray}

The equation of motion of each element of the Green's function $\textbf{G}$
can be obtained by \cite{Zubarev}
\begin{eqnarray}
\omega \langle\langle A \mid B \rangle\rangle
  =\langle [A, B]_{+}\rangle
  +\langle\langle [A, H] \mid B \rangle\rangle.
\label{Eq:ABEOM}
\end{eqnarray}
Applying the Hamiltonian of Eq.~(\ref{Eq_Ham}) to the above
equation, we obtain the equation of motion of
$G_{ij\sigma}^{\alpha\beta}$ as
\begin{eqnarray}
(\omega +\mu_\alpha) &\langle\langle& c_{i\alpha\sigma}
 \mid c_{j\beta\sigma}^\dag\rangle\rangle
  =\delta_{ij}\delta_{\alpha\beta}
     -\sum_{b\in Fe,m}T_{ib\alpha m}\langle\langle
     c_{bm\sigma}\mid c_{j\beta\sigma}^\dag\rangle\rangle\nonumber\\
  &&-\sum_{b\in Cu,\gamma}t_{ib}'\langle\langle c_{b\gamma\sigma}
     \mid c_{j\beta\sigma}^\dag\rangle\rangle
    +U\langle\langle n_{i\alpha\bar{\sigma}} c_{i\alpha\sigma}
     \mid c_{j\beta\sigma}^\dag\rangle\rangle\nonumber\\
  &&+U'\sum_{l\neq \alpha}\langle\langle
      n_{il\bar{\sigma}}c_{i\alpha\sigma}\mid
     c_{j\beta\sigma}^\dag\rangle\rangle
  +(U'-J)\sum_{l\neq \alpha}\langle\langle n_{il\sigma}
      c_{i\alpha\sigma}\mid c_{j\beta\sigma}^\dag\rangle\rangle,
\label{Eq:FOEOM}
\end{eqnarray}
where the second-order Green's functions, such as $\langle\langle
n_{il\bar{\sigma}} c_{i\alpha\sigma}\mid
c_{j\beta\sigma}^\dag\rangle\rangle$, appear on the right-hand side
of Eq.~(\ref{Eq:FOEOM}). Further, we construct the equations of
motion of all the second-order Green's functions that appear in
Eq.~(\ref{Eq:FOEOM}), and the higher-order Green's functions should
emerge accordingly. To obtain a self-consistent loop for the
calculation of the first-order Green's functions, some kinds of
decoupling approximations should be introduced. Based on the
Hubbard-I approximation \cite{Hubbard-I} for the one-band Hubbard
model, we develop a new decoupling scheme for the multi-orbital
systems, such as
\begin{eqnarray}
 \langle\langle n_{i\alpha\sigma'}c_{bm\sigma}\mid
    c_{j\beta\sigma}^\dag\rangle\rangle&\approx& \langle
    n_{i\alpha\sigma'}\rangle\langle\langle c_{bm\sigma}\mid
    c_{j\beta\sigma}^\dag\rangle\rangle,
    \nonumber\\
 \langle\langle c_{bm\sigma}^\dag c_{il\sigma}c_{i\alpha\sigma}\mid
    c_{j\beta\sigma}^\dag\rangle\rangle&\approx& \langle
    c_{bm\sigma}^\dag c_{il\sigma}\rangle\langle\langle
    c_{i\alpha\sigma}\mid c_{j\beta\sigma}^\dag\rangle\rangle\nonumber\\
  &&-\langle c_{bm\sigma}^\dag c_{i\alpha\sigma}\rangle\langle\langle
    c_{il\sigma}\mid c_{j\beta\sigma}^\dag\rangle\rangle,
   \nonumber\\
 \langle\langle n_{is\sigma'}n_{il\sigma''}c_{i\alpha\sigma}\mid
    c_{j\beta\sigma}^\dag\rangle\rangle&\approx& \langle
    n_{is\sigma'}\rangle\langle\langle
    n_{il\sigma''}c_{i\alpha\sigma}\mid
    c_{j\beta\sigma}^\dag\rangle\rangle\nonumber\\
  &&+\langle n_{il\sigma''} \rangle\langle\langle
    n_{is\sigma'}c_{i\alpha\sigma}\mid
    c_{j\beta\sigma}^\dag\rangle\rangle.
    \nonumber\\
\end{eqnarray}
As a result, we can approximately express the higher-order Green's
functions by the first-order Green's functions, making up a
self-consistent loop for the calculation of the inhomogeneous
Green's function $\textbf{G}$
\begin{equation}
\textbf{M}\cdot \textbf{G} = \textbf{N},
\end{equation}
where $\textbf{M}$ and $\textbf{N}$ are the same order matrices as
\textbf{G}, which can be obtained by
\begin{eqnarray}
\langle A B\rangle
=-\frac{1}{\pi}\int_{-\infty}^{+\infty}
f(\omega)\textmd{Im}\langle\langle A\mid B\rangle\rangle.
\end{eqnarray}
By solving the above self-consistent equations, we obtain the
real-space Green's function for a certain disordered configuration.
From which we can also obtain the local density of states (LDOS) at
site $i$ for orbital $\alpha$ by
$\rho_{\alpha}(\textbf{r}_i,\omega)=-\frac{1}{\pi}\textmd{Im}
G_{ii}^{\alpha\alpha}(\omega)$. To find the averaged DOS of the
whole system, we need to calculate the LDOS of different disorder
configurations, and then determine the sample-averaged values
afterwards.

The DOS of Cu and Fe ions can be expressed respectively as
\begin{eqnarray}
\rho_{\alpha}^{(Fe)}&=&\frac{1}{N_s}\sum_{s=1}^{N_s}\frac{1}{N_{Fe}}\sum_{i\in Fe}
\rho_{\alpha}^{(s)}(\textbf{r}_i,\omega)\nonumber\\
\rho_{\alpha}^{(Cu)}&=&\frac{1}{N_s}\sum_{s=1}^{N_s}\frac{1}{N_{Cu}}\sum_{i\in Cu}
\rho_{\alpha}^{(s)}(\textbf{r}_i,\omega),
\end{eqnarray}
where $N_{Fe}$ and $N_{Cu}$ represent the total numbers of the Fe
and Cu ions, respectively. $N_s$ is the number of disorder
configurations. The real-space Green's function method has many
advantages, such as simple in principle and suitable for many kinds
of correlated compounds with disorder. In addition, we can perform
self-consistent calculations for lattices with up to $32\times 32$
sites.

\section{Anderson Localization and Orbital Selectivity}
\label{Sec:U+J}

Our method not only handles the lattice disorder precisely,
but also treats the correlation more accurate than the mean-field
approximation, making it a priority to deal with the correlated
system with the presence of disorder. In this section, we use the
newly developed real-space Green's function method to study the
cooperative effect of multiorbital correlation and impurity-induced
disorder on the OSMT.

\subsection{OSMT tuned by Hund's rule coupling}
\label{Sec:J-undop}

\begin{figure}[t]
\begin{center} \includegraphics[width=0.7\columnwidth,clip]{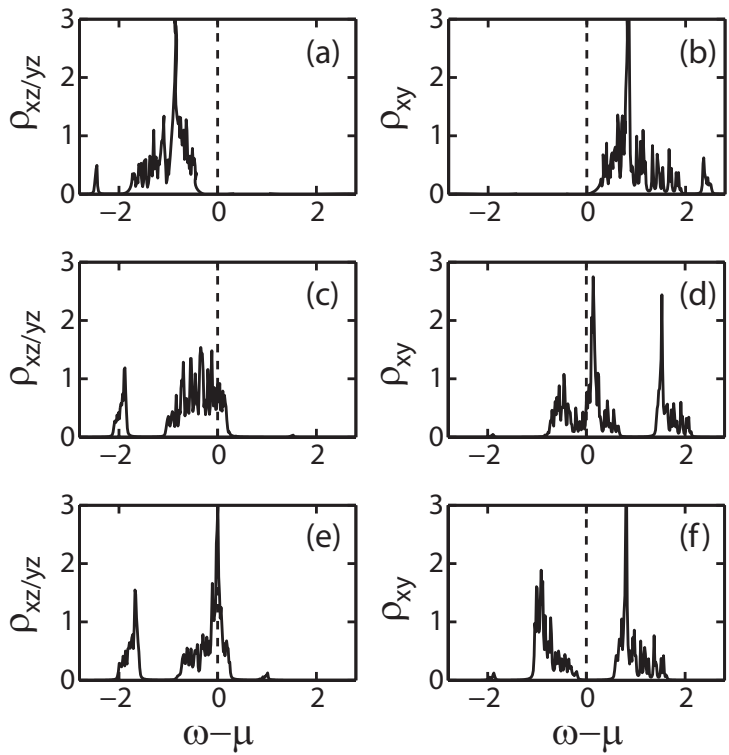} \end{center}
\caption{Density of states of $d_{xz}/d_{yz}$ orbitals (left panel)
and $d_{xy}$ orbital (right panel) for the undoped cases with
different Hund's rule couplings: $J$=0.05 eV (top panel), $J$=0.18
eV (middle panel), and $J$=0.21 eV (bottom panel). Top panel: a band
insulator with fully filled $d_{xz}/d_{yz}$ orbitals but empty
$d_{xy}$ orbital when $J$=0.05 eV; Middle panel: a metallic phase
when $J$=0.18 eV, where all three orbitals are metallic; Bottom
panel: an orbital-selective Mott phase when $J$=0.21 eV, where
$d_{xz}/d_{yz}$ orbitals are metallic, while $d_{xy}$ orbital
becomes a Mott insulator. The on-site correlation is $U$=1.5 eV, and
the relation $U=U'+2J$ is satisfied, as the same for the following
figures.} \label{fig:dosp}
\end{figure}

A clear way to show the effect of Hund's rule coupling on
the MIT is to omit the influence of the Cu substitution firstly. As
the undoped Fe-based superconductors are indicated to have a filling
of roughly two thirds based on the band structure calculations
\cite{Haule2009,Boeri2008}, we determine the Fermi energy by a
constraint of four electrons per Fe,
\begin{equation}
n_{total}=-\frac{1}{\pi}\sum_{i\alpha}\int_{-\infty}^{+\infty}
f(\omega)\textmd{Im} G_{ii}^{\alpha\alpha}(\omega)d\omega\nonumber=4,
\end{equation}
where $f(\omega)$ is the Fermi-Dirac distribution function. It was
predicted that, only at half-filling, the Hund's rule coupling $J$
is capable of strengthening the effective correlations to introduce a
Mott-Hubbard MIT \cite{Mecici2011,Han,Koga,Pruschke}. In the undoped
Fe-based superconductors, further research is needed to understand the
effect of the Hund's rule coupling on the OSMT in a three-orbital
system away from half-filling.

We study the influence of the Hund's rule coupling on the orbitally
resolved DOS in undoped Fe-based superconductors, and the results
are shown in Fig.~\ref{fig:dosp}. Consistent with the previous
study \cite{Medici2009}, it is shown that the OSMT is manipulated by
the Hund's rule coupling in the undoped Fe-based superconductors.
As we can see in Fig.~\ref{fig:dosp} (a) and (b), when the Hund's rule
coupling is very small ($J$=0.05 eV), the system becomes a band insulator
with the fully-filled $d_{xz}$ and $d_{yz}$ orbitals but the empty
$d_{xy}$ orbital, indicating that electrons prefer to occupy the
orbitals of lower energy. With the increase of the Hund's rule coupling,
all three orbitals become metallic shown in Fig.~\ref{fig:dosp} (c) and
(d) when $J$=0.18 eV, suggesting that electrons transfer from $d_{xz}$
and $d_{yz}$ orbitals to the empty $d_{xy}$ orbital. Once $J$ exceeds the
critical value $J_c\approx 0.2$ eV, the Mott-Hubbard MIT happens in the
$d_{xy}$ orbital, while the $d_{xz}$/$d_{yz}$ orbitals remain metallic,
leading to an OSMT in the undoped Fe-based superconductor.

\begin{figure}[t]
\begin{center} \includegraphics[width=0.7\columnwidth,clip]{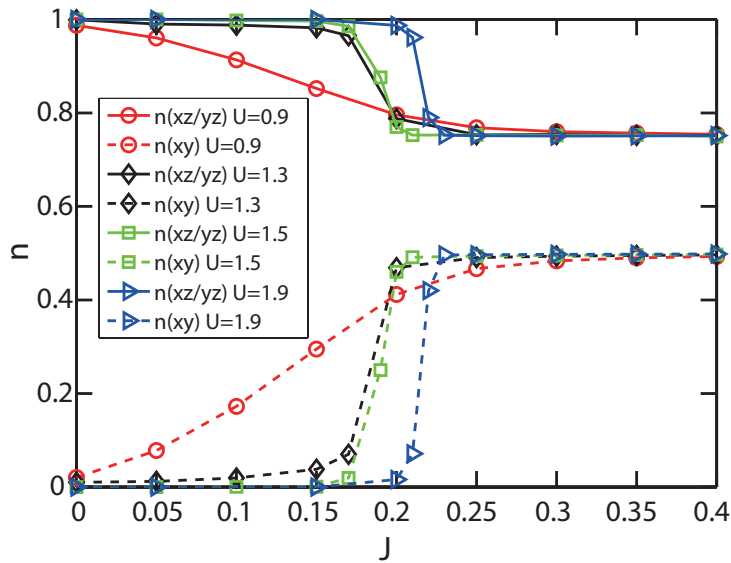} \end{center}
\caption{(Color online) Dependence of orbital occupancy on
Hund's rule coupling $J$ for undoped compounds with different
interactions $U$. For very small $J$, it is a band insulator
with fully-filled $d_{xz}/d_{yz}$ orbitals but empty $d_{xy}$
orbital. With the increase of the Hund's rule coupling, the
difference of orbital occupancies becomes weak, and the
$d_{xy}$ orbital tends to be half-filled when $J$ is larger than 0.2
eV, regardless of the value of $U$.} \label{fig:nj}
\end{figure}

It is obvious that the electron transition from the $d_{xz}/d_{yz}$
orbitals to $d_{xy}$ orbital plays an essential role in finding the
OSMP. We plot the effect of the Hund's rule coupling on the electron
occupancy of each orbital in Fig.~\ref{fig:nj}, where the crystal-field
splitting is fixed as $\Delta_{xy}$=0.75 eV according to the local
density calculations \cite{Yangliu,Daghofer2012}. As a result, the
$d_{xy}$ orbital locates much higher than the $d_{xz}/d_{yz}$ orbitals,
which leads to an extremely strong difference of orbital occupancies.
However, the Hund's rule coupling favors orbital compensation,
i.e., tends to equalize the different orbital populations. The OSMT
appears as a result of the influence of the strong Hund's rule coupling
$J$ overcoming the effect of the orbital level splitting $\triangle_{xy}$ 
\cite{Werner2007,Okamoto}. As shown in Fig.~\ref{fig:nj}, the increase
of Hund's rule coupling will decrease the difference of orbital
occupancies generated by the orbital level splitting. The most remarkable
portion is in the vicinity of $J$=0.2 eV, where the orbital occupancies
change significantly, especially for the cases with large correlations $U$.
When $J$ is larger than 0.25 eV, the orbital occupancies tend to be stable
and the $d_{xy}$ orbital becomes half-filled.

In undoped Fe-based superconductors, the increase of Hund's rule
coupling benefits the OSMT mainly in two aspects. First, the
increase of Hund's rule coupling will transfer more electrons from
the $d_{xz}/d_{yz}$ orbitals to the $d_{xy}$ orbital, causing the
$d_{xy}$ orbital half-filled and the $d_{xz}/d_{yz}$ orbitals
away from half-filled when $J>$0.2 eV.
Secondly, the Hund's rule coupling will enhance the effective
correlations in the half-filled $d_{xy}$ orbital when $J>$0.2 eV.
Therefore, the Mott-Hubbard MIT can happen only in the half-filled
$d_{xy}$ orbital, while the $d_{xz}/d_{yz}$ orbitals are still metallic.

\subsection{Cooperative effect of Hund's rule coupling and impurity-induced disorder}

\begin{figure}[t]
\begin{center} \includegraphics[width=0.75\columnwidth,clip]{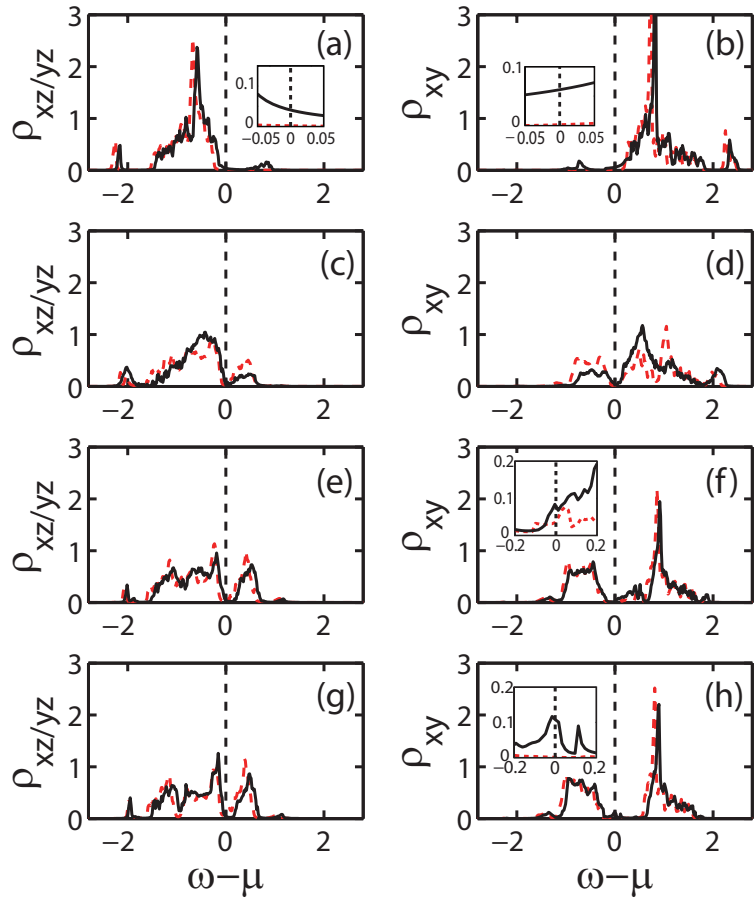} \end{center}
\caption{(Color online) Doping dependence of orbitally resolved
density of states for different impurity concentrations:
$x$=0.04 (dashed line) and $x$=0.12 (solid line). Top panel: a doping
induced band insulator to Anderson insulator transition when
$J$=0.05 eV; Second panel: Efros-shklovskii gap survives the strong
impurity concentration when $J$=0.18 eV; Third panel: the influence
of Cu substitution on the orbital selective Mott phase when $J$=0.21
eV, where the in-gap states are accumulated inside the Mott gap of the
$d_{xy}$ orbital; Bottom panel: an orbital-selective insulating
phase appears in the lower doping case x=0.04 when $J$=0.25 eV, where the
$d_{xz}/d_{yz}$ orbitals are Anderson localized but the $d_{xy}$ orbital
is a Mott insulator. The system transfer to Anderson
insulator with the the increase of doping concentration. Insets in
(a), (b), (f), and (h): enlargement of the orbitally resolved
density of states near the Fermi level. The correlations are $U$=1.5
eV and $u$=3 eV, and the results are averaged by 100 samples
respectively.} \label{fig:dosh}
\end{figure}

In this subsection, we study the effect of Cu substitution on the OSMT
introduced by the Hund's rule coupling. In the Fe-based superconductors
with the presence of Cu substitution, the electron occupations of
Fe $3d$ orbitals will be manifestly affected by the impurities. In addition,
it is found that impurity scattering introduced by Cu substitution has a
strong effect on the electron transport, which can lead to Anderson MIT in
iron-based compound. In Fig.~\ref{fig:dosh}, we present the effect of Cu substitution on
the OSMT by comparing the orbitally resolved DOS for different doping
concentrations ($x=0.04$ and $x=0.12$) with various Hund's couplings.
Firstly, with the increase of Cu impurities, a slight broadening of the
bandwidths is found for the band insulating phase with $J=0.05$ eV. As a
result, a small amount of electron states can be found across the Fermi
level when $x=0.12$ as shown by the insets in Fig.~\ref{fig:dosh}(a) and (b).
On the contrary, we find that the Cu impurities can induce a pseudogap at
the Fermi level for all three orbitals when $J=0.18$ eV, as shown in
Fig.~\ref{fig:dosh}(c) and (d).

The strong suppression on the DOS around
Fermi level derives from the cooperative effect of multiorbital
interactions and impurity-induced disorder. Efros and Shklovskii
demonstrated that the interactions between the localized electrons in a
disordered system can create a Coulomb gap in the DOS near the Fermi level
\cite{Efros}. Therefore, the appearance of the zero-bias anomaly at the
Fermi energy is an evidence for the Anderson localization of
electronic states in the correlated systems with disorder \cite{Song}.
As some experiments show, as well as the density functional
calculations demonstrate, some transition-metal substitutions can introduce
localization effect in iron-based superconductors\cite{Wadati,YangH2013,DengQ},
which is consistent with our findings. Furthermore, the pseudogap in the
$d_{xz}/d_{yz}$ orbitals can still be found for the case with $J=0.21$ eV
(Fig.~\ref{fig:dosh}(e)) or $J=0.25$ eV (Fig.~\ref{fig:dosh}(g)),
which indicates that it is robust against the increasing Hund's rule
coupling. In addition, the pseudogap will also affect the spectrum functions
and Fermi surfaces significantly.

\begin{figure}[t]
\begin{center} \includegraphics[width=0.95\columnwidth,clip]{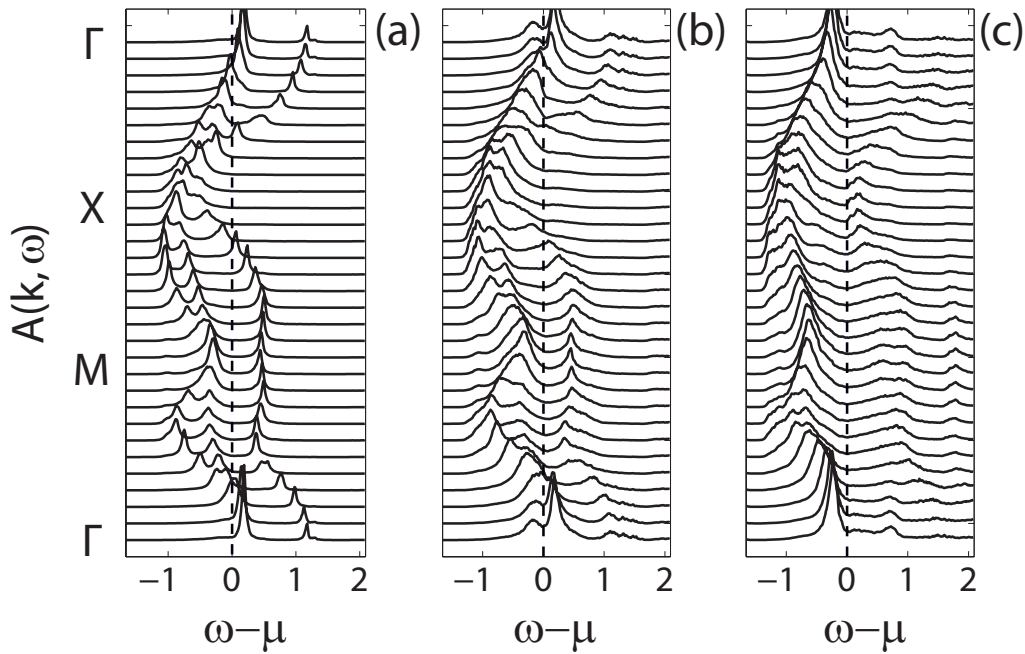} \end{center}
\caption{Spectrum functions for different substitution concentrations:
$x$=0.01 (a), $x$=0.04 (b), and $x$=0.08 (c) when $J$=0.18 eV.
The Fermi topology changes significantly with the the
increasing Cu concentration, and all Fermi pockets are pushed away from
the Fermi level when $x$=0.08. The correlation parameters are $U$=1.5 eV
and $u$=3 eV. Each result is obtained by averaging 100 disorder samples.}
\label{fig:specx}
\end{figure}

In Fig.~\ref{fig:specx}, we plot our result
of the impurity effect on the spectrum functions for the cases with fixed
Hund's rule coupling $J$=0.18 eV. Comparing with the undoped cases, we find an
apparent broadening of the orbital bandwidths, especially for the $d_{xy}$
orbital, in agreement with the experimental findings. This spectra-broadening
behavior induced by the substitution has been experimentally observed in
LiFe$_{1-x}$Co$_x$As \cite{BroadLFCA}, NaFe$_{1-x}$Co$_x$As \cite{BroadNFCA},
and Fe$_{1.04}$Te$_{1-x}$Se \cite{BroadFTS}, which is believed to be an
impurity-scattering effect induced by the dopants \cite{Broaden}. Once the
dopants deviate from the Fe-plane, as the case of Ba$_{1-x}$K$_x$Fe$_2$As$_2$,
the broadening effect disappears \cite{BroadBKFA}. With the increase of doping
concentration, we find that the top of band for the inner hole pocket shifts
downward below the Fermi level, leading to the disappearance of the inner hole
pocket, as shown in Fig.~\ref{fig:specx}(b). It is worth noticing that this is
just the embodiment of the appearence of the Efros-shklovskii gap at the Fermi
surface. If the doping concentration is continuously increased, all bands will
be pushed away from the Fermi level, leading all Fermi pockets vanish completely.

\begin{figure}[t]
\begin{center} \includegraphics[width=0.75\columnwidth,clip]{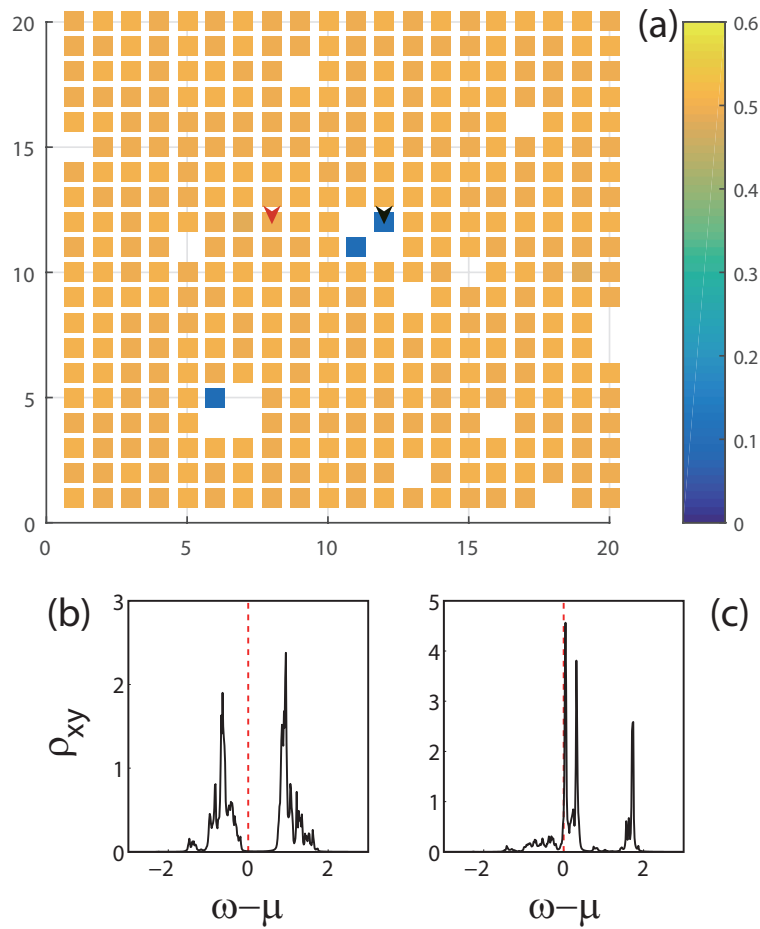} \end{center}
\caption{(Color online) The real-space distribution of the electron
occupancy on $d_{xy}$ orbital of iron atoms in a 20 $\times$ 20 lattice
with 16 randomly selected Cu substitutions shown by the lattice vacancies.
The value of $n_{xy}$ decreases significantly to 0.1 at some specific sites
which are very close to the highly concentrated area of the impurities.
These specific sites aside the other sites remain to be half-filling.
The LDOS of two representative sites, which are marked by red and black arrows,
are shown in Fig (b) and (c) respectively. The correlation parameters are $U$=1.5 eV, $J$=0.21 eV, and $u$=3 eV.}
\label{fig:ndis}
\end{figure}

As for the OSMP introduced by the Hund's rule coupling in undoped Fe-based
superconductors, the effect of Cu substitution will be mainly reflected
in two aspects. One is to induce a pseudogap in the $d_{xz}/d_{yz}$ orbitals,
as shown in Fig.~\ref{fig:dosh}(e) and (g).
The other effect is the appearance of the in-gap states
in the $d_{xy}$ orbital. Comparing Fig.~\ref{fig:dosh}(f) with Fig.~\ref{fig:dosp}(f), it is shown that, in the presence of Cu substitution, some electronic states will emerge
inside the Mott gap of the $d_{xy}$ orbital around the Fermi level.
It is worth noting that these in-gap states are all localized,
which can be confirmed directly from its origin.
Through observing the real-space fluctuation
of charge density induced by the substitution of copper ions, we find that the
occupancy of the $d_{xy}$ orbital deviates from half-filling at some discrete iron sites,
which are indicated by blue (dark) squares in Fig.~\ref{fig:ndis} (a).
Obviously, these peculiar sites are easier to be found in the highly concentrated area of
the impurities, where the real-space fluctuations induced by Cu substitution are much stronger.
As a result, LDOS of these specific lattice points will be greatly changed.
As shown in Fig.~\ref{fig:ndis} (c) for the LDOS of a site with $n\approx0.1$,
it is found that the Mott gap is replaced by a sharp peak near the Fermi level,
suggesting that the
in-gap states in the DOS of the $d_{xy}$ orbital are heavily localized.
In contrast, the iron ions at most other sites will keep the $d_{xy}$
orbital half-filled, where a Mott gap can be found at the Fermi level
for the LDOS of the $d_{xy}$ orbital, as shown in Fig.~\ref{fig:ndis} (b)
for the LDOS of a representative iron site.

\begin{figure}[t]
\begin{center} \includegraphics[width=0.95\columnwidth,clip]{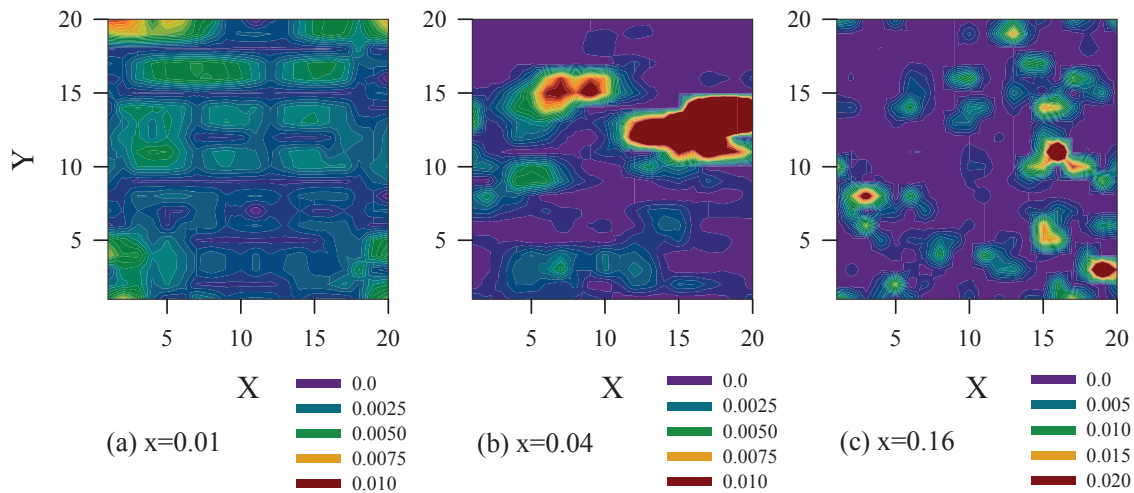} \end{center}
\caption{(Color online) For a certain disorder configuration of a
20$\times$20 lattice, the real-space distribution of the LDOS is
shown for an electronic state with energy very close to the Fermi
level when the impurity concentration increases from $x=0.01$ (a) to
$x=0.04$ (b), and to $x=0.16$ (c). The localization of the
electronic state is enhanced significantly with the increasing
doping concentration. The correlation parameters are $U$=1.5 eV,
$J$=0.18 eV, and $u$=3 eV.} \label{fig:Ldos}
\end{figure}

To visually demonstrate the substitution-induced localization for all three
orbitals, we show a real-space distribution of the LDOS of an electronic state
with the energy very close to the Fermi level in Fig.~\ref{fig:Ldos} for
a certain substitution configuration.
In the case of a light doping ($x=0.01$), it is shown that the LDOS distributes
homogenously in the real space,
suggesting that the state is weakly localized.
With the increase of the impurity concentration,
the LDOS tends to be restricted in some small puddles, where the sizes
of the puddles decrease significantly when the impurity concentration
increases from $x=0.04$ (Fig.~\ref{fig:Ldos}(b)) to $x=0.16$
(Fig.~\ref{fig:Ldos}(c)).  It is demonstrated that very strong Anderson
localization has been introduced by the Cu substitutions in the cases of heavy
doping.

In order to preciously determine the critical point of Anderson MIT,
we perform the lattice-size scaling of the conductance $g$, which is
obtained by a simplified form of Kubo formula in the case with
$\omega=0$ \cite{Kubo,PALee},
\begin{eqnarray}
g&=&\frac{e^2}{\pi} \textmd{Tr}
         [\textmd{Im}G_i(j,j')\textmd{Im}G_i(j'-1,j-1)
         +\textmd{Im}G_i(j-1,j'-1)\textmd{Im}G_i(j',j)\nonumber\\
      &&-\textmd{Im}G_i(j,j'-1)\textmd{Im}G_i(j',j-1)
         -\textmd{Im}G_i(j-1,j')\textmd{Im}G_i(j'-1,j)],\nonumber\\
\end{eqnarray}
where the trace is over the sites perpendicular to the current
direction ($k$ direction). ($i$, $j$) denote the coordinates of
lattice sites in the ribbon, where the periodic boundary conditions
should be applied only to the perpendicular direction to the
current. $j$ and $j'$ are chosen to be on opposite sides of the
lattice along the current direction, respectively. The conductance
of the whole system is obtained by summarizing the three diagonal
elements of the conductance matrix $g$, corresponding to the
conductances of all three orbitals.

\begin{figure}[t]
\begin{center} \includegraphics[width=0.75\columnwidth,clip]{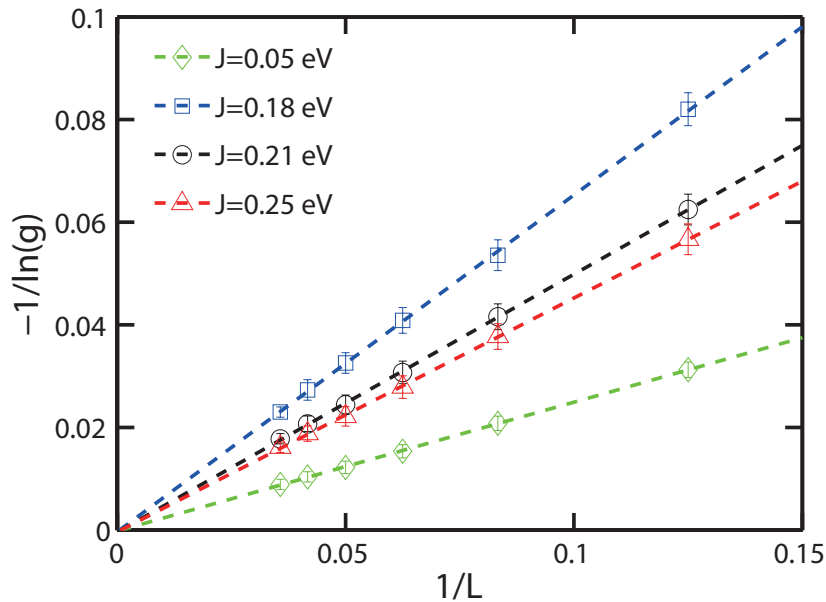} \end{center}
\caption{(Color online) Lattice-size scaling of conductance $g$:
$-1/ln(g)$ vs $1/L$ for different Hund's rule coupling.
The linear fittings indicate that the
conductance should decay exponentially with the lattice size.
The doping concentration is
$x$=0.01, and one hundred disordered samples are considered.}
\label{fig:Scale}
\end{figure}

The conductance is expected to has the form
$g\propto e^{-2L/\lambda}$
in an Anderson localized system \cite{Kramer}, where $\lambda$ is
the localization length and $L$ denotes the lattice size. In
Fig.~\ref{fig:Scale} we present the fitting of the lattice-size
scaling of conductance of the weakly doped system ($x=0.01$) for the
cases with different Hund's rule coupling. We find linear
relationship of $-1/ln(g)$ vs $1/L$, suggesting that
the system is Anderson localized for all cases with different Hund's
rule coupling. In addition, no evidence of a
significant delocalization effect of the Hund's rule coupling has
been found, while a non-monotonic behavior is presented for the
effect of $J$ on the Anderson localization shown by the scaling of
the conductance $g$. It is shown that the localization length
decrease monotonously when the Hund's rule coupling increases
form $J$=0.18 eV to $J$=0.25 eV. On the one hand, the Hund's rule
coupling can lead to Mott-Hubbard MIT in the $d_{xy}$ orbital.
On the other hand, the Hund's rule coupling
also has strong effect to suppress the in-gap states around the
Fermi level. As a result, the contribution of the $d_{xy}$ orbital
to the conductance will decrease with the increasing Hund's rule
coupling. However, the situation is completely different when
$J<$0.1 eV. As shown in Fig.~\ref{fig:dosp}(a) and (b),
the system behaves as a band insulator when $J<$0.1 eV. Therefore,
the conductance takes an extremely small value owing to the lack
of states at the Fermi level. With the increase of the Hund's rule
coupling, more and more electrons are transferred from the $d_{xz}/d_{yz}$
orbitals to the $d_{xy}$ orbital as shown in Fig.~\ref{fig:nj}, accompanied
with the increase of electronic states at the Fermi level for all three
orbitals. Therefore, the conductance increases firstly with the increasing
Hund's rule coupling. While when the OSMT happens at about $J$=0.21 eV,
a decrease of $g$ can be found owing to enhancement of the suppression on
the in-gap states with the increasing Hund's rule coupling.

\begin{figure}[t]
\begin{center} \includegraphics[width=0.75\columnwidth,clip]{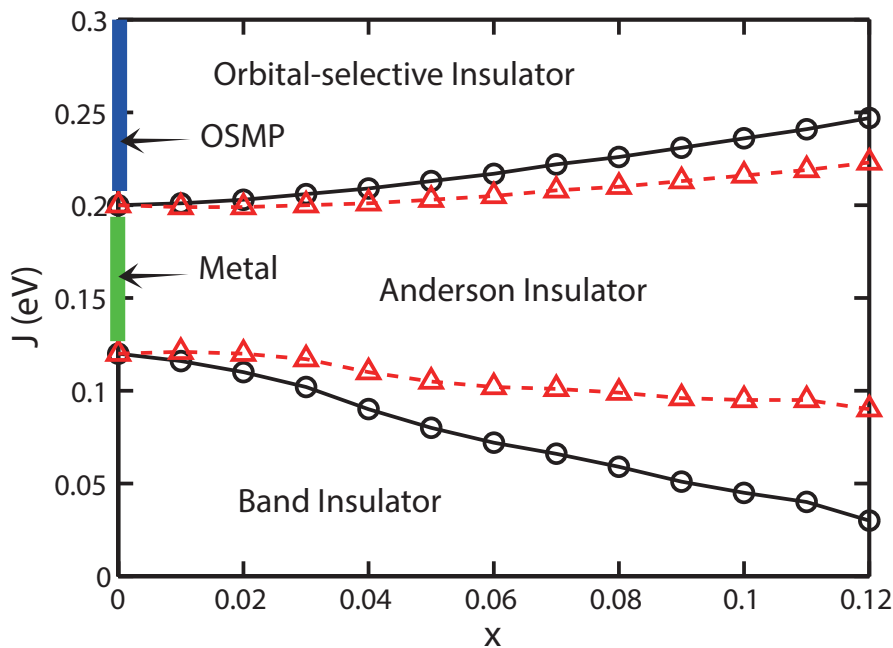} \end{center}
\caption{(Color online) Phase diagram of Cu-substituted iron-based
superconductors. The solid lines with circles (dashed lines with
triangles) indicate the phase boundaries of hole doping cases with
$u$=3 eV (electron doping cases with $u$=6 eV).} \label{fig:phas}
\end{figure}

We compile all our finding about the cooperative effect of the
Hund's rule coupling and the doping-induced disorder on the
MIT in a complete $J-x$ phase diagram,
as shown in Fig.~\ref{fig:phas}. An OSMP, where a Mott gap
opens in the $d_{xy}$ orbital but the degenerate $d_{xz}/d_{yz}$
orbitals are still metallic, has been found in the undoped case when
the Hund's coupling is $J$=0.21 eV.
In the presence of Cu substitution,
the Mott gap of the $d_{xy}$ orbital is fullfilled by the in-gap states
when $x$ increases to 0.12, as shown in Fig.~\ref{fig:dosh}(f). However,
If the Hund's rule coupling increases to $J$=0.25 eV, the doping induced
in-gap states will be suppressed, even completely disappear in the case
with $x$=0.04, as shown by the inset in Fig.~\ref{fig:dosh}(h). Therefore,
we find an orbital-selective insulating phase under the impact of both
disordered substitution and multi-orbital correlation, where the $d_{xz}/d_{yz}$
orbitals are Anderson localized while the $d_{xy}$ orbital is a Mott insulator.
In addition, as indicated in Fig.~\ref{fig:dosp}(f) and Fig.~\ref{fig:dosh}(f),
with the increase of the doping concentration, the whole
multiorbital system will transfer from an orbital-selective insulator
to an Anderson insulator when $J\geq$ 0.21 eV. It is worth noticing that
no metallic phase can be found for the Fe-based superconductor with the
presence of Cu substitution. Based on the scaling theory \cite{Patrick},
all states should be localized for arbitrary weak disorder in the
two-dimensional systems. Our phase is consistent with this conclusion.

In our study, we find that when the
on-site interactions $u$ of Cu ions is larger than the critical
value 5.2 eV, the hole doping will be converted into the electron
doping \cite{Yangliu}. However, the effects of
Cu substitution on the OSMT are qualitatively the same for both the
hole and electron dopants. Therefore, our study is mainly concentrated
in the hope doping cases with $u$=3 eV.  Besides, the carrier density
will also be changed with the increasing Cu substitution.
As a result, the area of the band
insulator exhibits continued shrinkage with the increasing doping
concentration, especially for the hole doping cases. The reason
why the effect of the hole doping is more obvious than that of the
electron doping is because the real-space fluctuation of the orbital
occupancy is stronger in the hole doping cases.
On the other hand, to suppress the in-gap states introduced
by the Cu impurities, we should enhance the Hund's rule coupling to
maintain the Mott gap. As a result, the boundary between the
orbital-selective and Anderson insulating phases
gradually moves upwards, owing to an increasing of the critical
$J_c$ for the Mott-Hubbard MIT transition in the $d_{xy}$ orbital.

\section{Conclusion}
\label{Sec:Con}

We study the cooperative effect of the Hund's rule coupling and the
doping-induced disorder in iron-based superconductors with Cu
substitution. The Hund's rule coupling benefits the
orbital-selective Mott transition by transferring electrons from
$d_{xz}/d_{yz}$ orbitals to $d_{xy}$, and enhancing the correlation
in the half-filled $d_{xy}$ orbital as well. The Anderson
localization introduced by the impurities is robust against
variations in the Hund's rule coupling. A whole phase diagram
depended on both the Hund's rule coupling and the doping
concentration is obtained, where the Cu substitution affects the
orbital-selective Mott phase by introducing localized in-gap states.
As a result of the Anderson localization induced by the impurities, an
orbital-selective insulating phase is found for the Cu-substituted
Fe-based superconductors, where the $d_{xy}$ orbital are Mott
insulator while the degenerate $d_{xz}/d_{yz}$ orbitals are Anderson
localized.

\section*{Acknowledgments}
We thank Prof. Liang-Jian Zou and  Da-Yong Liu for helpful
discussion. The computational resources utilized in this research
were provided by Shanghai Supercomputer Center. The work was
supported by National Natural Science Foundation of China
(Nos. 10974018, 11174036 and 11474023), the National Basic Research 
Program of China (No. 2011CBA00108), and the Fundamental Research Funds for the
Central Universities.


\section*{References}
\begin{thebibliography}{100}



\bibitem{Lebegue2007}
Lebegue S 2007 Phys. Rev.B 75 035110

\bibitem{Singh2008}
Singh D J and Du M H 2008 Phys. Rev. Lett.100 237003

\bibitem{Hosono2006}
Kamihara Y, Hiramatsu H, Hirano M, Kawamura R, Yanagi H,
Kamiya T and Hosono H 2006 J. Am. Chem. Soc. {\bf 128} 10012

\bibitem{Hirschfeld2011}
Hirschfeld P J, Korshunov M M and Mazin I I 2011 Rep. Prog. Phys. {\bf
74} 124508

\bibitem{ChenX2014}
Chen X, Dai P, Feng D, Xiang T and Zhang F C 2014 Nat. Sci. Rev. {\bf 1} 371

\bibitem{Haule2009}
Haule K, Kotliar G 2009 New J. Phys. {\bf 11} 025021

\bibitem{Mecici2011}
de¡¯Medici L 2011 Phys. Rev. B {\bf 83} 205112

\bibitem{Medici2014}
deMedici L, Giovannetti G and Capone M 2014 Phys. Rev. Lett. {\bf 112} 177001

\bibitem{Fanfarillo2015}
 Fanfarillo L and Bascones E 2015 Phys. Rev. B {\bf 92} 075136

\bibitem{YuR2013}
Yu R, Si Q 2013 Phys. Rev. Lett {\bf 110} 146402

\bibitem{YiM2015}
Yi M et al.2015 Nat. Commun.6:7777 doi: 10.1038/ncomms8777

\bibitem{YiM2013}
Yi M, Lu D H, Yu R, Riggs S C, Chu J H, Lv B, Liu Z K, Lu M, Cui Y T,
Hashimoto M, Mo S K,Hussain Z, Chu C W, Fisher I R, Si Q and Shen Z X 2013
Phys. Rev. Lett. 110 067003

\bibitem{WangZ2014}
Wang Z, Schmidt M, Fischer J, Tsurkan V, Greger M, Vollhardt D, Loidl A
and Deisenhofer J 2014 Nat. Commun. 5:3202 doi:10.1038/ncomms4202

\bibitem{DingXX2014}
Ding X X, Pan Y M, Yang H, Wen H H 2014 Phys. Rev. B {\bf 89} 224515

\bibitem{Rev-DMFT}
Georges A, Kotliar G, Krauth W and Rozenberg M J 1996 Rev. Mod.
Phys. {\bf 68} 13


\bibitem{Anisimov2002}
Anisimov V I, Nekrasov I A, Kondakov D E, Rice T M and Sigrist M 2002
Eur. Phys. J. B {\bf 25} 191

\bibitem{Werner2007}
Werner P, Millis A J 2007 Phys. Rev. Lett. {\bf 99} 126405

\bibitem{Medici2009}
de¡¯Medici L, Hassan S R, Capone M and Dai X 2009
Phys. Rev. Lett. {\bf 102} 126401

\bibitem{SongZ2014}
Song Z Y, Lee H, and Zhang Y Z 2015 New J. Phys. {\bf 17} 033034

\bibitem{McLeod}
McLeod J A, Buling A, Green R J, Boyko T D, Skorikov N A,
Kurmaev E Z, Neumann M, Finkelstein L D, Ni N, Thaler A,
Bud\rq{}ko S L, Canfield P C and Moewes A 2012 J. Phys: Condens. Matter
{\bf 24} 215501

\bibitem{YanYJ}
Yan Y J, Cheng P, Ying J J, Luo X G, Chen F, Zou H Y, Wang A F,
Ye G J, Xiang Z J, Ma J Q and Chen X H 2013 Phys. Rev. B {\bf 87} 075105

\bibitem{ChengP}
Cheng P, Shen B, Han F and Wen H H 2013 Eur. Phys.
Lett. {\bf 104} 37007

\bibitem{Ideta}
Ideta S, Yoshida T, Nishi I, Fujimori A, Kotani Y, Ono K, Nakashima Y,
Yamaichi S, Sasagawa T, Nakajima M, Kihou K, Tomioka Y,
Lee C H, Iyo A, Eisaki H, Ito T, Uchida S and Arita R 2013
Phys. Rev. Lett. {\bf 110} 107007

\bibitem{Merz}
Merz M, Schweiss P, Nagel P, Wolf Th, L\"{o}hneysen H v and Schuppler S
2016 J. Phys. Soc. Jpn. {\bf 85} 044707

\bibitem{WangAF}
Wang A F, Lin J J, Cheng P, Ye G J, Chen F, Ma J Q, Lu X F,
Lei B, Luo X G and Chen X H 2013 Phys. Rev. B {\bf 88} 094516

\bibitem{CuiST}
Cui S T, Kong S, Ju S L, Wu P, Wang A F, Luo X G,
Chen X H, Zhang G B and Sun Z 2013 Phys. Rev. B {\bf 88} 245112

\bibitem{HuangTW}
Huang T W, Chen T K, Yeh K W, Ke C T, Chen C L,
Huang Y L, Hsu F C, Wu M K, Wu P M, Avdeev M and Studer A J 2010
Phys. Rev. B {\bf 82} 104502
%

\bibitem{NiN}
Ni N, Thaler A, Yan J Q, Kracher A, Colombier E,
Bud\rq{}ko S L and Canfield P C 2010 Phys. Rev. B {\bf 82} 024519

\bibitem{LiJ}
Li J, Guo Y F, Zhang S B, Yuan J, Tsujimoto Y, Wang X,
Sathish C I, Sun Y, Yu S, Yi W, Yamaura K, Takayama-Muromachiu E,
Shirako Y, Akaogi M and Kontani H 2012 Phys. Rev. B {\bf 85} 214509

\bibitem{Williams}
Williams A J, McQueen T M, Ksenofontov V, Felser C and Cava R J 2009
J. Phys.: Condens. Matter {\bf 21} 305701


\bibitem{AL50years}
Byczuk K, Hofstter W and Vollhard D 2010
Chapter 20 {\it 50 years of Anderson Localization} ed. E. Abrahams
(World Scientific)

\bibitem{Hubbard-I}
Hubbard J 1963 Proc. Roy. Soc. A {\bf 276} 238

\bibitem{Yangliu}
Liu Y,Liu D Y, Wang J L, Sun J, Song Y and Zou L J 2015
Phys. Rev. B {\bf 92} 155146

\bibitem{Daghofer2010}
Daghofer M, Nicholson A, Moreo A and Dagotto E 2010 Phys. Rev. B
{\bf 81} 014511

\bibitem{Daghofer2012}
Daghofer M, Nicholson A and Moreo A 2012 Phys. Rev. B {\bf 85} 184515

\bibitem{YiM2012}
Yi M, Lu D H, Moore R G, Kihou K, Lee C H, Iyo A,
Eisaki H, Yoshida T, Fujimori A and Shen Z X 2012
New J. Phys. {\bf 14} 073019

\bibitem{Zubarev}
Zubarev D N 1960 Soviet Phys. Usp. (English Trasl.) {\bf 3} 320

%
%


\bibitem{Boeri2008}
Boeri L, Dolgov O V and Golubov A A 2008
Phys. Rev. Lett. {\bf 101} 026403





\bibitem{Han}
Han J E, Jarrell M and Cox D L 1998 Phys. Rev. B {\bf 58} R4199

\bibitem{Koga}
Koga A, Imai Y and Kawakami N 2002 Phys. Rev. B {\bf 66} 165107

\bibitem{Pruschke}
Pruschke T and Bulla R 2005 Eur. Phys. J. B {\bf 44} 217

\bibitem{Okamoto}
Okamoto S, Millis A J 2004 Phys. Rev. B {\bf 70} 195120

\bibitem{Efros}
Efros A L and Shklovskii B I 1975 J. Phys. C: Solid State Phys.
{\bf 8} L49

\bibitem{Song}
Song Y, Bulut S, Wortis R and Atkinson W A 2009
J. Phys.: Condens. Matter {\bf 21} 385601

\bibitem{Wadati}
Wadati H, Elfimov I and Sawatzky G A 2010
Phys. Rev. Lett. {\bf 105} 157004

\bibitem{YangH2013}
Yang H, Wang Z Y, Fang D L, Deng Q, Wang Q H, Xiang Y Y,
Yang Y, Wen H H 2013 Nat. Commun. 4:2749 doi: 10.1038/ncomms3749

\bibitem{DengQ}
Deng Q, Ding X X, Li S, Tao J, Yang H and Wen H H 2014
New J. Phys. {\bf 16} 063020

\bibitem{BroadLFCA}
Pitcher M J, Lancaster T, Wright J D, Franke I, Steele A J,
Baker P J, Pratt F L, Thomas W T, Parker D R,
Blundell S J and Clarke S J 2010 J. Am. Chem. Soc. {\bf 132} 10467

\bibitem{BroadNFCA}
Parker D R, Smith M J P, Lancaster T, Steele A J,
Franke I, Baker P J, Pratt F L, Pitcher M J, Blundell S J
and Clarke S J 2010 Phys. Rev. Lett. {\bf 104} 057007

\bibitem{BroadFTS}
Chen F, Zhou B, Zhang Y, Wei J, Ou H W, Zhao J F,
He C, Ge Q Q, Arita M, Shimada K, Namatame H,
Taniguchi M, Lu Z Y, Hu J, Cui X Y and Feng D L 2010
Phys. Rev. B {\bf 81} 014526

\bibitem{Broaden}
Ye Z R, Zhang Y, Chen F, Xu M, Jiang J, Niu X H, Wen C H P, Xing L Y,
Wang X C, Jin C Q, Xie B P and Feng D L 2014
Phys. Rev. X {\bf 4} 031041

\bibitem{BroadBKFA}
Chen H, Ren Y, Qiu Y, Bao W, Liu R H, Wu G, Wu T,
Xie Y L, Wang X F, Huang Q and  Chen X H 2009
Europhys. Lett. {\bf 85} 17006






\bibitem{Kubo}
Kubo B R 1957 J. Phys. Soc. Jpn. {\bf 12} 570
%
\bibitem{PALee}
Lee P A and Fisher D S 1981 Phys. Rev. Lett. {\bf 47} 882

\bibitem{Kramer}
Kramer B and MacKinnon A 1993 Rep. Prog. Phys. {\bf 56} 1469

\bibitem{Patrick}
Lee P A and Ramakrishnan T V 1985 Rev. Mod. Phys. {\bf 57} 287

\end{thebibliography}
\end{document}